\documentclass[runningheads]{llncs}
\usepackage[T1]{fontenc}
\usepackage{graphicx}
\usepackage{hyperref}
\usepackage{xcolor}
\usepackage{xspace}
\usepackage{listings}
\usepackage{orcidlink}
\usepackage{tabularx}
\usepackage{booktabs}
\usepackage{censor}

\definecolor{jsonkey}{rgb}{0.8, 0.2, 0.2}     
\definecolor{jsonvalue}{rgb}{0.2, 0.4, 1.0}   
\definecolor{jsonnumber}{rgb}{0.741,0.333,0.196} 
\definecolor{jsoncomment}{rgb}{0.609,0.617,0.499} %
\definecolor{jsonmeta}{rgb}{0.215,0.231,0.266} 

\lstdefinelanguage{json}{
    basicstyle=\ttfamily\scriptsize,
    numbers=left,
    numberstyle=\tiny\color{gray},
    stepnumber=1,
    numbersep=5pt,
    showstringspaces=false,
    breaklines=true,
    frame=tb,
    backgroundcolor=\color{jsoncomment!20},
    comment=[l]{\%},
    morecomment=[l][\color{jsoncomment}]{\%},
    commentstyle=\color{jsoncomment},
    moredelim=**[is][\color{jsonkey}]{@}{@},
    morestring=[b]",
    stringstyle=\color{jsonvalue},
    literate=
     *{:}{{{\color{jsonmeta}:}}}{1}
      {,}{{{\color{jsonmeta},}}}{1}
      {\{}{{{\color{jsonmeta}{\{}}}}{1}
      {\}}{{{\color{jsonmeta}{\}}}}}{1}
      {[}{{{\color{jsonmeta}[}}}{1}
      {]}{{{\color{jsonmeta}]}}}{1}
      {0}{{{\color{jsonnumber}0}}}{1}
      {1}{{{\color{jsonnumber}1}}}{1}
      {2}{{{\color{jsonnumber}2}}}{1}
      {3}{{{\color{jsonnumber}3}}}{1}
      {4}{{{\color{jsonnumber}4}}}{1}
      {5}{{{\color{jsonnumber}5}}}{1}
      {6}{{{\color{jsonnumber}6}}}{1}
      {7}{{{\color{jsonnumber}7}}}{1}
      {8}{{{\color{jsonnumber}8}}}{1}
      {9}{{{\color{jsonnumber}9}}}{1}
}

\newcommand{\one}{({\textit{i}})\xspace}
\newcommand{\two}{({\textit{ii}})\xspace}
\newcommand{\three}{({\textit{iii}})\xspace}
\newcommand{\four}{({\textit{iv}})\xspace}
\newcommand{\five}{({\textit{v}})\xspace}

\begin{document}
\title{Towards Multi-Agent Economies: Enhancing the A2A Protocol with Ledger-Anchored Identities and x402 Micropayments for AI Agents}
\titlerunning{Towards Multi-Agent Economies}
%

\author{Awid Vaziry\inst{1,2}\orcidlink{0009-0007-2192-5968} \and
Sandro Rodriguez Garzon\inst{1,2}\orcidlink{0000-0001-6921-294X} \and
Axel Küpper\inst{1,2}\orcidlink{0000-0002-4356-5613}}
\authorrunning{A. Vaziry et al.}

\institute{Technische Universität Berlin,
  Straße des 17. Juni 135, 10623 Berlin, Germany
\email{\{vaziry,sandro.rodriguezgarzon,axel.kuepper\}@tu-berlin.de}\\
T-Labs, Ernst-Reuter-Platz 1, 10587 Berlin, Germany}

\maketitle              
\begin{abstract}
This research article presents a novel architecture to empower multi-agent economies by addressing two critical limitations of the emerging Agent2Agent (A2A) communication protocol: decentralized agent discoverability and agent-to-agent micropayments. By integrating distributed ledger technology (DLT), this architecture enables tamper-proof, on-chain publishing of AgentCards as smart contracts, providing secure and verifiable agent identities. The architecture further extends A2A with the x402 open standard, facilitating blockchain-agnostic, HTTP-based micropayments via the HTTP 402 status code. This enables autonomous agents to seamlessly discover, authenticate, and compensate each other across organizational boundaries. This work further presents a comprehensive technical implementation and evaluation, demonstrating the feasibility of DLT-based agent discovery and micropayments. The proposed approach lays the groundwork for secure, scalable, and economically viable multi-agent ecosystems, advancing the field of agentic AI toward trusted, autonomous economic interactions.

\end{abstract}
%
%
%
\section{Introduction}
\label{sec:introduction}
Agentic AI is an emerging paradigm for process and task automation based on large language models (LLMs), which can autonomously pursue long-term goals, make decisions, and execute complex, multi-turn workflows. Multi-agent systems (MAS) require the collaboration and data exchange among agents operating across different security domains. Future use cases, such as agent-based travel booking, may require API calls to various services or cross-domain collaboration among AI agents. Therefore, agents must be able to discover other agents, evaluate their trustworthiness, and compensate other agents for their services using micropayments~\cite{birch_agentic_2025,rothschild_agentic_2025,sapkota_ai_2025}.
The Agent2Agent\footnote{https://a2aproject.github.io} (A2A) protocol was introduced by Google in April 2025, supported by numerous industry partners. It establishes an industry standard for discovering agent capabilities, which are advertised at the agent's endpoint through the AgentCard, and facilitates agent-to-agent communication within and across security domains. However, the protocol lacks two crucial features to enable a multi-agent economy: \one a trusted database for agent discovery and reputation management that allows agents to find other agents and evaluate their trustworthiness and reliability before engaging in transactions, and \two payment functionality to compensate other agents for their services.


This article presents a novel architecture that facilitates agentic AI collaboration through distributed ledger technology (DLT)-based agent discovery and DLT-based micropayments, leveraging HTTP 402 and utilizing HTTP headers and status code 402 to enable micropayments. The proposed architecture introduces two key novelties: \one blockchain-based AgentCards that allow agents to advertise their capabilities on-chain; and \two an HTTP-based micropayment flow for the A2A protocol.

By combining the A2A protocol for inter-agent communication with DLT for AgentCard discovery and the x402~\cite{reppel_x402_2025} open standard HTTP 402 implementation for DLT-agnostic micropayments, our architecture enables autonomous agents to collaborate across organizational boundaries. This work advances the field of agentic AI by offering a practical architecture for secure, trustworthy, and economically viable multi-agent economies.


\section{Background}
\label{sec:background}
This section lays the groundwork for understanding how technologies from the fields of AI agent communication, DLT-based identity and micropayments interact and complement each other. The discussion begins with the fundamentals of LLMs and AI agents, with a focus on emerging standards for agent interoperability and collaboration. The application of DLT in multi-agent collaboration scenarios is then explored. Finally, micropayment protocols that could facilitate economic interactions between autonomous agents are investigated.

\subsection{Large Language Models and AI Agents}
LLMs are predominantly transformer-based neural language models pre-trained on extensive text corpora, utilizing a self-attention mechanism~\cite{vaswani_attention_2017}. Transformers are effective at tasks such as text generation, translation, summarization, and sentiment analysis~\cite{minaee_large_2025}.

An AI agent is an autonomous system capable of perceiving its environment, reasoning about its objectives, and executing actions to achieve specific goals, often utilizing LLMs for advanced decision-making and planning. Agentic AI extends this concept to ecosystems of such agents that interact, collaborate, and coordinate with one another and with humans to solve complex tasks. These multi-agent systems are distinguished by their ability to delegate subtasks to specialized agents with complementary capabilities, adapt dynamically, and operate across organizational boundaries. As AI agents become essential to intelligent applications, establishing clear interface definitions, standardized protocols, interoperability frameworks, and robust governance structures becomes necessary to ensure their reliable and secure deployment~\cite{yang_survey_2025,habler_building_2025,sanabria_beyond_2025}.


\subsection{Agent2Agent Protocol}
The Agent2Agent protocol is an open protocol that facilitates communication and collaboration between AI agents. By providing a common framework, it allows agents created with different technologies and by various vendors to work together seamlessly. The protocol defines the "Client Agent" and the "Server/Remote Agent". Remote Agents perform actions based on received tasks from the Client Agent and return results through the A2A protocol. Agents promote their capabilities through metadata files at their endpoints, using a standardized JSON format file known as the AgentCard\footnote{According to the specification advertised at: https://<host>/.well-known/agent.json}. Communication is structured around tasks with defined lifecycles and outputs, supporting asynchronous, multimodal, and authenticated exchanges. A2A is primarily designed to enable scalable collaboration among diverse agents spanning various platforms and vendors.~\cite{rao_surapaneni_agent2agent_2025}. The protocol claims to facilitate discoverability through the use of the AgentCard. However, the A2A specifications do not specify how agents can discover other AgentCards. The challenge of agent discovery, particularly across diverse security domains and organizational boundaries, has yet to be fully addressed\footnote{https://a2aproject.github.io/A2A/v0.2.5/topics/agent-discovery/}$^{,}$\footnote{https://github.com/a2aproject/A2A/discussions/741}. 


\subsection{Distributed Ledger Technology and Blockchain}
Distributed Ledger Technology (DLT) enables the maintenance of append-only transactional databases in a decentralized manner. Blockchain, the most prominent DLT implementation, secures data by bundling transactions into cryptographically linked blocks. This concept was introduced for the Bitcoin cryptocurrency~\cite{nakamoto_bitcoin_2008}, and extended by Ethereum with smart contracts for programmable transactions. An Ethereum smart contract is a program deployed to the blockchain, written in high-level programming languages such as Solidity. Smart contracts on a public network can be accessed and interacted with by anyone. However, the functions of the contract that allow writing data can be restricted programmatically to specific accounts, groups, or other smart contracts. When deployed on public networks, contracts follow the rules and economic models of those networks, which usually include transaction fees for computation and storage to prevent spam and compensate node operators. These fees, denominated in units such as “gas” in Ethereum, create resource constraints that impact performance~\cite{androulaki_hyperledger_2018}. As blockchain technology has evolved, it has led to innovations such as stablecoins, which are on-chain tokens that aim to maintain a stable value by pegging to an external asset, often a fiat currency such as the US dollar~\cite{jajodia_encyclopedia_2025}.

DLT offers several key advantages that can be leveraged in the context of MAS. First, a ledger is a distributed store of value and data that is usually tamper-proof. Permissionless networks can be utilized as a public database for advertising the capabilities of agents without a single entity in control, and can also implement additional features, such as a rating system or other identity verification features, for these agents. Another key advantage of DLTs is that they enable the existence and exchange of cryptocurrencies, providing an alternative to traditional currencies and payment processors such as banks, credit card companies, or networks such as PayPal. Transaction fees on blockchains vary between different networks and may sometimes be higher than those charged by traditional systems. However, depending on the specific ledger used, these fees can also be significantly lower than those of conventional alternatives. These low fees and fast direct settlements, without intermediaries or complex onboarding procedures, make micropayments economically and technologically viable. For AI agents, DLT thus offers not only auditability through immutable record-keeping and secure information sharing but also discoverability, token-based payment, and incentive mechanisms that can be leveraged to enable multi-agent economies.

\subsection{HTTP Status Code 402: Payment Required}
The HTTP 402 "Payment Required" status code is a standard client error response indicating that the requested content is inaccessible until a payment is made. The code is marked as reserved for future use, meaning most browsers or web services do not commonly use it at this time\footnote{https://datatracker.ietf.org/doc/html/rfc7231\#section-6.5.2}. However, recent initiatives utilize the HTTP 402 status code to facilitate web-based microtransactions involving various currencies. The following presents two implementations of HTTP 402. The first is one of the earliest initiatives to utilize DLT-based payments for accessing resources over HTTP, the L402 protocol. The second is one of the most recent implementations, the x402 protocol.

\subsubsection{L402: Lightning HTTP 402 Protocol}
The Lightning HTTP 402 protocol (L402) builds on Macaroon credentials and the Bitcoin Lightning Network. Initially introduced for cloud services, macaroons are similar to web cookies but encode permissions and support decentralized authorization. They allow for embedded conditions (caveats) that define when a target service should authorize requests~\cite{birgisson_macaroons_2014}. The Bitcoin Lightning Network is a second-layer technology built on top of the Bitcoin blockchain designed to enable faster and cheaper transactions. It allows users to create off-chain payment channels where multiple transactions can occur instantly before the final balance is settled on the main Bitcoin blockchain~\cite{poon_bitcoin_2016}.
\subsubsection{x402 Protocol}
\begin{figure}[h!]
    \centering
    \includegraphics[width=\textwidth]{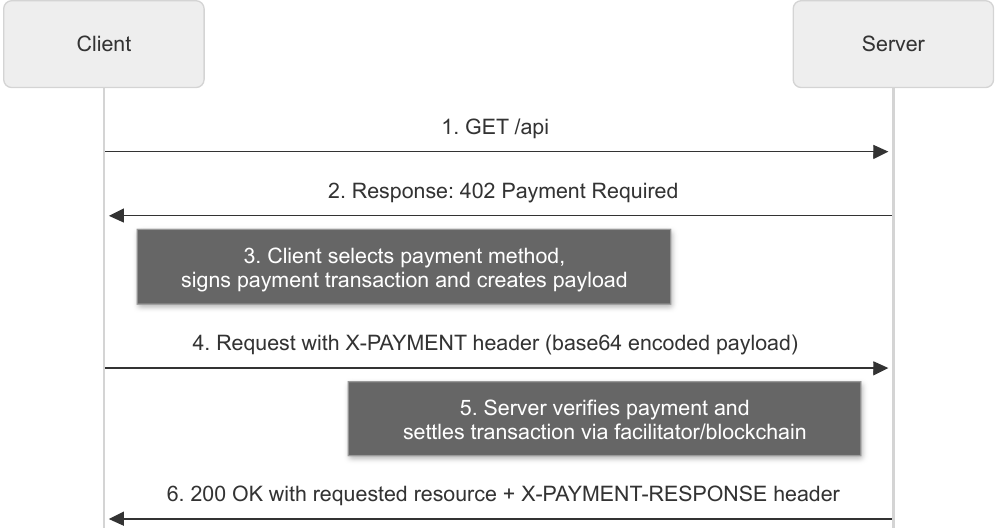}
    \caption{Simplified x402 payment flow with initial payment required (HTTP 402) response and successful payment.}
    \label{fig:x402-sequence}
\end{figure}
The x402 protocol~\cite{reppel_x402_2025} is an open protocol introduced and maintained by Coinbase Global, Inc., based on a concept similar to that of L402, although it is agnostic to both tokens and blockchains. This means the payment receiver (payee) can choose the asset and the blockchain network they wish to use. As indicated in \autoref{fig:x402-sequence}, when an x402-enabled API is accessed without an attached payment, it returns a 402 response containing the necessary metadata for the payment. This metadata includes information such as the payment token address, the receiver contract address, the blockchain network, the amount to be paid, the timestamp, and a nonce. 
The payer then prepares and signs a payment transaction, which they include in the header of their next API call, base64-encoded in the "X-PAYMENT" field. The payee verifies the transaction and sends it to an endpoint of a blockchain node for on-chain settlement. If the transaction succeeds, the API responds appropriately to the payer and returns the "X-PAYMENT-RESPONSE" in the header~\cite{reppel_x402_2025}.

In contrast to L402, designed to utilize the Lightning Network ecosystem, the x402 protocol offers a more flexible and generalized approach to web-based micropayments. While L402 utilizes the Lightning Network for off-chain settlement and employs Macaroons for granular, decentralized authorization, x402 is blockchain- and token-agnostic. The x402 implementation is primarily realized through HTTP headers and supports a wide range of cryptocurrencies and DLT networks, with on-chain settlement. The architectural differences extend to the authorization model:  In L402, Macaroons are utilized to embed authorization conditions, whereas in x402, a valid signed payment transaction serves as the authorization for the request. This makes x402 a simpler and adaptable solution for developers aiming to integrate micropayments across diverse blockchain ecosystems, whereas L402 is a more specialized protocol for the Bitcoin and Lightning Network environment.                 
\section{Related Work}
\label{sec:related-work}
The related work section focuses on two areas of research related to agentic AI and MAS. First, existing studies on micropayments that occur both between agents and between agents and human users are examined. Second, research on establishing the discoverability of the A2A AgentCard is explored.

The authors in~\cite{rothschild_agentic_2025} predict a shift from the current model of isolated service agents and end-to-end agents to systems that enable seamless communication between agents. This evolution is anticipated to follow two main pathways: closed "agentic walled gardens" controlled by dominant platforms or an open "web of agents" in which agents are able to connect and transact freely. This transformation is expected to drive a shift towards the use of micropayment transactions, as traditional payment methods are considered unsuitable due to the high overhead costs associated with payments. Achieving this vision, especially the open "web of agents", is expected to require a robust infrastructure for agent discovery, identity verification, and efficient micropayments.

The ability for AI agents to conduct payments and manage identities has recently emerged as a rapidly evolving area of research. While an increasing amount of literature discusses the potential of agent-to-agent micropayments~\cite{birch_agentic_2025}, both peer-reviewed academic studies and grey literature presenting and implementing solutions on this topic remain limited. The available solutions and frameworks are primarily driven by enterprises implementing payment solutions\footnote {https://paymanai.com/} and micropayments\footnote{https://www.mastercard.com/news/press/2025/april/mastercard-unveils-agent-pay-pioneering-agentic-payments-technology-to-power-commerce-in-the-age-of-ai/}$^{,}$\footnote{https://skyfire.xyz/} between AI agents. 


The discoverability of agents in the A2A ecosystem is primarily facilitated by AgentCards, which serve as the protocol-level artifact for exposing agent metadata at endpoints. The current A2A specification describes three main discovery mechanisms: \one direct retrieval via a well-known URI at \texttt{/.well-known/
agent.json}, \two curated registries in which agents self-register and are discoverable by metadata queries, and \three explicit configurations for closed or private networks\footnote{https://a2aproject.github.io/A2A/latest/specification/\#5-agent-discovery-the-agent-card}$^{,}$\footnote{https://a2aproject.github.io/A2A/topics/agent-discovery/}.

A recent comprehensive analysis by Ehtesham et al.~\cite{ehtesham_survey_2025} positions the AgentCard approach in A2A as state of the art for trusted enterprise task delegation within organizational boundaries, but identifies a critical limitation: A2A assumes agent catalog availability and is fundamentally enterprise-centric, thereby rendering it inapplicable to most open internet scenarios. This assessment is consistent with the warning of Li et al., who warn that tightly coupling discovery and execution can increase security risks when separation layers are unclear~\cite{li_glue-code_2025}. The survey reveals that while A2A excels at inter-agent negotiation and artifact-driven delegation, it does not address the broader challenge of cross-platform agent discovery across untrusted domains~\cite{ehtesham_survey_2025}.
The limitations of A2A become especially apparent when considering alternative solutions, such as the Agent Network Protocol (ANP). While the existing approaches rely heavily on a registry-dependent model for agent discovery, the ANP aims to overcome these constraints by utilizing W3C Decentralized Identifiers (DIDs) for open-network agent discovery. 
However, ANP faces practical deployment challenges, including high negotiation overhead and the need for a robust ecosystem to facilitate adoption.~\cite{ehtesham_survey_2025}. Community recognition of discovery limitations of A2A is reflected in GitHub proposals for multi-agent indices (e.g., \texttt{/.well-known/agents.json}) and API-catalog schemas, indicating demand for scalable enumeration and cross-domain discovery\footnote{https://github.com/a2aproject/A2A/discussions/166}$^{,}$\footnote{https://github.com/a2aproject/A2A/pull/642}. The Agent Name Service (ANS) proposal by Huang et al.~\cite{huang_agent_2025} attempts to bridge this gap through a DNS-inspired, protocol-agnostic registry infrastructure, but relies on centralized PKI and governance structures, trading decentralization for verifiable identity and lifecycle management.

The literature suggests that current A2A discovery mechanisms represent an intermediate solution rather than the final architecture for agent interoperability. ANP's capabilities in DID-based trustless identity and AI-native protocol negotiation highlight the technical path forward for addressing A2A's cross-domain discovery limitations. Despite these advances, a critical gap persists: the absence of a unified approach that combines A2A's enterprise-grade AgentCard model with ANP's decentralized discovery capabilities. Current solutions force a binary choice between enterprise-focused registry systems and fully decentralized, yet complex, DID-based approaches. This leaves the unresolved challenge of seamless agent discovery across organizational boundaries while maintaining the security and reliability guarantees required for autonomous economic transactions. Addressing this gap is essential for realizing the broader vision of a multi-agent economy where autonomous agents can discover, authenticate, and pay other agents within and across trust domains.                 
\section{System Architecture}
\label{sec:system-architecture}
This section presents the proposed architecture for the system, which enables decentralized discovery and micropayments for AI agents. The architecture introduces two core innovations: \one decentralized, DLT-based publishing of agent identity cards, \two seamless agent-to-agent micropayments. Current challenges in agentic web scenarios include discovering available agents, verifying their skills, endpoints, and payment channels, and binding these elements in a trustworthy manner. To address these issues, this work proposes embedding agent cards within blockchain-based smart contracts, creating a decentralized registry that offers secure binding of agent endpoints and capabilities, as well as dynamic update mechanisms controlled by the agent owner. The proposed design capitalizes on several fundamental properties of DLT. Specifically, DLT permits any participant to create new entries, guarantees universal visibility of all published records, and restricts modification rights. Furthermore, the continuous synchronization of the ledger across all network nodes ensures that the registry remains consistent and tamper-evident. The distributed ledger acts as a single source of truth for agent identity cards, ensuring that no participant can alter or duplicate entries belonging to others, while still allowing everyone to read the entries. Additionally, the token functionality of the ledger is utilized to integrate token-based micropayments into agent smart contracts. This architecture provides built-in payment addresses and a verifiable payment history, enabling the development of robust metrics for agent reputation. For instance, the more often agents pay for a service from a particular agent, the more trustworthy that agent becomes. The following sections describe the architecture for decentralized and discoverable AgentCard publishing, utilizing DLT and DLT-based micropayments through the x402 protocol for the A2A protocol.

\begin{lstlisting}[language=json, caption={Mandatory fields of the AgentCard agent.json}, label={lst:agentcard-json}]
{
  "name": "", %Name of the Agent (Human readable)
  "description": "", %Agent Description (Human readable)
  "url": "", %Base URL for A2A. Must be HTTPS
  "version": "1.0.0", %Agent version
  "capabilities": {}, %Optional A2A feeatures, e.g. streaming
  "defaultInputModes": ["text/plain"], %Accepted input media types
  "defaultOutputModes": ["application/json"], %Produced output media types
  "skills": [ %Array of skills.
    {
      "id": "",
      "name": "",
      "description": ""
    }
  ]
}
\end{lstlisting}

\subsection{Decentralized, DLT-based, AgentCard Publishing}
The AgentCard is a JSON data structure that provides a standardized overview of the agent's features, including its name, description, version, data format, URL endpoint, and a detailed list of its specific skills and capabilities. \autoref{lst:agentcard-json} shows the mandatory fields of the AgentCard. The concept of on-chain agent smart contracts is introduced, referring to smart contracts deployed on public blockchains (such as Ethereum) that function as decentralized, publicly available identity cards for autonomous AI agents. These contracts encode agent identity, capabilities, endpoints, and other metadata in a standardized format, allowing them to advertise their services, skills, and payment details in a transparent and publicly accessible manner.

\subsubsection{Agent Smart Contract}
An agent smart contract includes all structured data fields defined in the AgentCard, along with a function that returns the \textit{agent.json}. Additionally, it incorporates economic functionalities, such as the ability to receive payments in native blockchain coins or approved ERC20 tokens. The contract provides secure management of withdrawal permissions and includes mechanisms for access control and ownership of the agent contract. By leveraging the immutability and openness of blockchain networks, agent contracts provide a trustless foundation for agent discovery, reputation building, and automated economic interactions. They enable agents to be discovered, authenticated, and compensated without reliance on centralized intermediaries. A contract variable indicating whether the agent remains active is further proposed. This variable is based on either transaction activity concerning the contract or a function requiring the agent to confirm its active status after a defined interval, or it involves both approaches. The reputation of an agent is derived either directly by implementing a rating mechanism into the contract, allowing other agents that have transacted with the agent's smart contract to rate the services. Alternatively, the reputation is indirectly inferred by examining the number and frequency of transactions and recurring transactions associated with the agent contract.

\subsubsection{Agent Discovery}
\label{sec:sub_agent-discovery}
\begin{table}[ht]
\centering
\caption{Agent Discoverability Methods: Comparative Analysis}
\label{tab:agent-discoverability}
\renewcommand{\arraystretch}{1.2}
\begin{tabularx}{\textwidth}{>{\raggedright\arraybackslash}p{3cm} >{\raggedright\arraybackslash}X >{\raggedright\arraybackslash}X >{\raggedright\arraybackslash}X}
\toprule
\textbf{Discovery by} & \textbf{Description} & \textbf{Advantages} & \textbf{Limitations} \\
\midrule
\textbf{Factory Contract} & 
Agents are deployed via a standardized factory contract & 
Knowing one contract makes all contracts from this factory discoverable& 
Limited to a single factory, reduced flexibility, vendor lock-in \\
\midrule
\textbf{On-Chain Registry} & 
Agents voluntarily enroll in an on-chain registry contract & 
Open participation, decentralized querying, no gatekeepers, or optional gatekeepers (e.g., consortium for quality control) & 
Self-registration required, potential spam, or centralization \\
\midrule
\textbf{Off-Chain Indexing} & 
Third-party indexers monitor and aggregate agent contracts automatically & 
High performance, rich filtering options, real-time updates & 
Infrastructure dependency, centralization risks, potential data lag \\
\midrule
\textbf{ERC Standard + Aggregators} & 
Agents implement common standards; specialized aggregators provide discovery & 
Ecosystem-wide reach, seamless integration, community-driven & 
Adoption timeline, standard fragmentation, implementation overhead \\
\bottomrule
\end{tabularx}
\end{table}
The discoverability of agents can be achieved in various ways. This section outlines four promising methods to discover other agent contracts. First, if the Solidity factory contract pattern~\cite{book:98402717} is utilized, contracts are derived from a so-called (master) factory smart contract, which saves costs and standardizes deployment. If one contract deployed by the factory is known, the factory can be queried to list all agent contracts it has created. This enables tracing any agent contract back to its factory and discovering all related agent contracts. The second discovery option is an on-chain registry contract that allows agents to enroll in it, which can be implemented in two variants: either in a fully decentralized manner, allowing any agent to enroll themselves, or alternatively, the list can be curated by an organization or consortium for quality control and adherence to regulations. The third option, off-chain indexing, involves third-party providers indexing all contracts that define an AgentCard, which could then be implemented as a free or paid service. Lastly, if a standard parallel to ERC-20 (Fungible Token Standard) or ERC-721 (Non-Fungible Token Standard) is established for on-chain agent identities, it would encourage data aggregators such as Etherscan, CoinMarketCap, and others to begin collecting information about agent contracts. This would lead to the creation of curated lists of these agents, similar to what is observed with specific tokens, such as ERC-20 and ERC-721.

\autoref{tab:agent-discoverability} illustrates the primary trade-offs associated with various agent discovery mechanisms in decentralized systems. The \textbf{factory contract pattern} approach provides a high degree of standardization and facilitates straightforward enumeration of agent deployments, thereby enhancing compatibility and auditability. However, this method’s dependence on a single factory contract introduces risks of vendor lock-in and restricts flexibility, potentially impeding the evolution of customized agent contracts and features as the ecosystem matures.

By contrast, the \textbf{on-chain registry} model fosters openness and decentralization, as it allows agents to enroll in a publicly accessible on-chain registry autonomously. These registries are fully decentralized and open, or they are curated registries of vetted agents. This approach either introduces central points of control or is susceptible to challenges such as spam registrations, which discourage participation or compromise the quality of the registry. Furthermore, the requirement for registration implies that agents not proactively enrolled remain undiscoverable through this mechanism.

The \textbf{off-chain indexing} paradigm relies on external entities, so called indexers, to monitor, collect, and aggregate agent contract data. This method enables advanced filtering, rapid updates, and scalable querying capabilities, all of which are essential for supporting extensive agent ecosystems. Nevertheless, this approach introduces a degree of centralization, as it necessitates trust in the reliability and neutrality of the indexing infrastructure. Additionally, potential delays in data synchronization and the risk of infrastructure failure may undermine the trustless guarantees that are foundational to blockchain-based architectures.

The adoption of an \textbf{ERC standard} for agent contracts, supplemented by aggregator services, represents an effective and standardisation-based avenue for achieving broad interoperability and seamless integration within the ecosystem. The success of standards such as ERC-20 and ERC-721 demonstrates the viability of this approach in promoting community-driven aggregation and cross-platform compatibility. The effectiveness of this model depends on widespread adoption. However, early-stage fragmentation and implementation complexities pose challenges as the standard develops. Furthermore, the standardization of on-chain agent identities has yet to be incubated, indicating that this approach will require time to develop.

In summary, the comparative analysis presented in the table demonstrates that each discovery mechanism has distinct advantages and limitations. This suggests that a composite strategy may offer the most effective and resilient framework for agent discovery in decentralized environments. Therefore, a mix of approaches could be adopted to leverage the strengths of each mechanism while mitigating their drawbacks.

\subsection{Agent2Agent Simple x402 Micropayment Flow}
\begin{figure}[ht!]
    \centering
    \includegraphics[width=\textwidth]{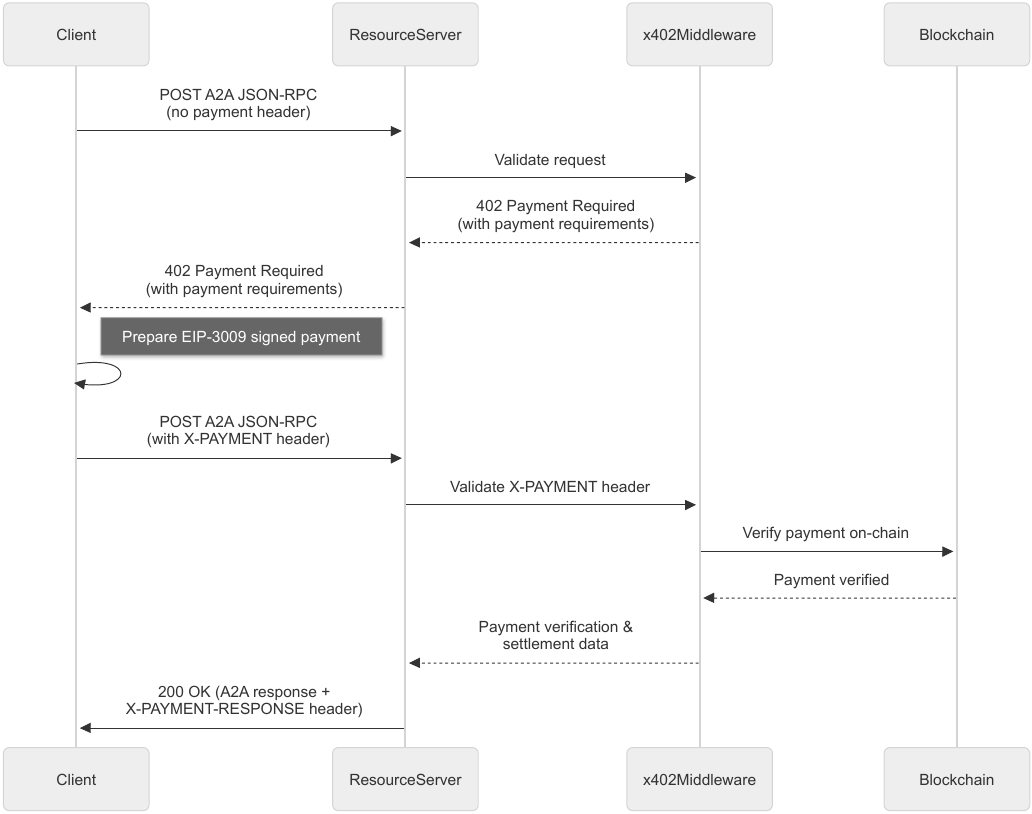}
    \caption{Simplified x402 integrated A2A message and payment flow with initial payment required (HTTP 402) response and successful payment.}
    \label{fig:x402+a2a-sequence}
\end{figure}

The proposed architecture integrates the x402 micropayment protocol with the Agent2Agent protocol, enabling seamless, blockchain-agnostic payments between autonomous agents. This section outlines the conceptual flow and core components necessary for secure, verifiable agent-to-agent micropayments, as depicted in the sequence diagram in \autoref{fig:x402+a2a-sequence}. The simplified x402-integrated A2A sequence consists of four main components: 
\begin{enumerate}
    \item A client that requests a resource.
    \item A resource server that offers the resource.
    \item The x402 middleware running on the resource server, which handles requests and processes payment-related operations to ensure the resource is protected from unpaid requests.
    \item A blockchain network used for payment settlement.
\end{enumerate}

The sequence is initiated when a client (e.g., Agent A) attempts to access a resource from the server (e.g., Agent B). If a request is made without a valid payment transaction in the HTTP header, the middleware responds with an HTTP 402 "Payment Required" status, including the relevant payment metadata.
This metadata encompasses the necessary transaction parameters, including the payment token address (defining the asset desired to be paid in), payment receiver address, blockchain network specification, payment amount, and other information.
Subsequently, the client prepares an EIP-3009 compliant, signed payment transaction, which is then embedded in the X-PAYMENT header field as a base64-encoded payload for the subsequent A2A JSON-RPC request. EIP-3009 is an Ethereum token standard that enables “transfer with authorization,” allowing the token owner to pre-authorize transfers by signing a cryptographically secure message, similar to writing a bank check to a defined recipient. This signed authorization can then be submitted on-chain by any party, separating payment authorization and execution. The x402 open standard also includes safeguards such as a unique nonce and a validity period to prevent replay attacks. By utilizing EIP-3009, the architecture enables agents to delegate payment execution to the recipient or a third party, ensuring trustless, verifiable, and efficient micropayment settlement before service delivery.
After receiving the base64-encoded transaction, the x402 middleware component validates the payment header and submits the transaction on-chain to the blockchain network.
Upon successful payment verification and settlement, the resource server responds with the requested A2A response, accompanied by an X-PAYMENT-RESPONSE header, thereby completing the transactional cycle.
Through this flow, this work combines the A2A protocol with the x402 micropayment protocol, allowing compensation for A2A resources.

\subsection{Architectural Summary}
The proposed architecture achieves seamless integration of the A2A and x402 protocols by leveraging the layered structure of HTTP communication. This design maintains a clear separation between the protocols while enabling them to work together effectively. Payment-related data is transmitted using dedicated HTTP headers: the X-PAYMENT header carries payment credentials within the A2A request, and the X-PAYMENT-RESPONSE header returns settlement details in the response. Meanwhile, the HTTP response body continues to use the standard JSON-RPC 2.0 format required by the A2A specification. Notably, when a client sends an initial request containing valid payment credentials, the interaction is fully compatible with both A2A and x402 protocols. This enables a seamless, standards-compliant transaction flow without deviation from the expected protocol behaviors.

A key feature of this architecture is that the DLT-based AgentCard can provide all necessary payment information, including recipient addresses, supported payment tokens, and the amount. This enables clients to construct and submit payment transactions proactively, without needing to query the agent for payment details beforehand. As a result, clients can initiate payment-enabled requests immediately, streamlining agent-to-agent interactions and reducing latency. There is one notable exception: if a client omits payment credentials, the server responds with an HTTP 402 "Payment Required" message, including the required payment metadata in the response body as specified by the x402 protocol. This temporarily departs from the standard A2A JSON-RPC format to ensure the client receives all information needed to complete the payment.

Once payment is successfully processed, the system returns to full compliance with the protocol. The response body resumes the standard JSON-RPC success format, while the X-PAYMENT-RESPONSE header provides payment settlement information as a base64-encoded payload. In summary, this architecture extends A2A functionality with micropayments in a way that is interoperable and non-disruptive, requiring no changes to the underlying protocol standards. 
\section{Implementation and Evaluation}
To validate the proposed architecture for integrating x402 micropayments with the A2A protocol, a comprehensive prototype system was developed, implementing the core components outlined in \autoref{sec:system-architecture}. The implementation serves as a proof-of-concept for DLT-based agent discovery and seamless micropayment integration within multi-agent economies.
\subsection{System Architecture Technical Implementation}
The prototype features a modular architecture consisting of five primary components: \one an A2A-compliant service agent, with integrated \two x402 payment middleware, \three a blockchain facilitator service for payment verification and settlement, \four a smart contract-based agent identity containing the AgentCard, and \five a mock USDC token contract, deployed on a local Hardhat development blockchain network. The system was also deployed and tested on the Sepolia testnet using the official USDC contract\footnote{Sepolia USDC contract: 0x1c7D4B196Cb0C7B01d743Fbc6116a902379C7238}. 

The prototype was developed using TypeScript, with Express.js for HTTP server functionality and ethers.js for blockchain interactions. This technology stack ensures compatibility with existing A2A systems and supports reliable integration with blockchain networks. The use of established frameworks also facilitates future system extension and maintenance.

\subsection{Agent Smart Contract Technical Implementation}
\begin{figure}
    \centering
    \includegraphics[width=\textwidth]{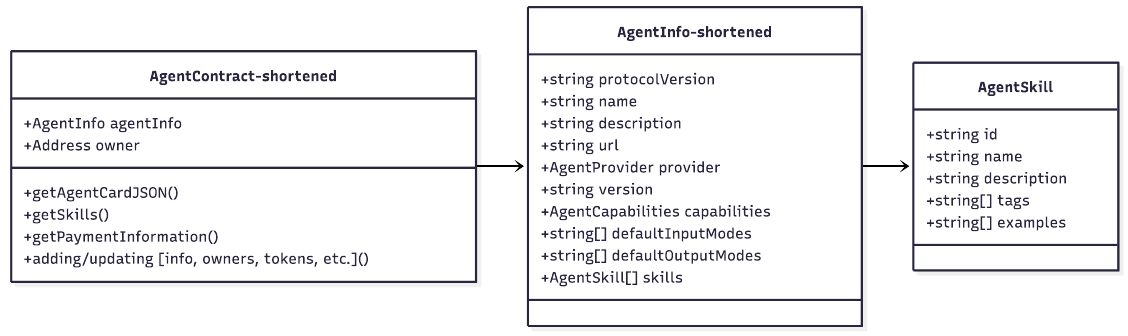}
    \caption{Illustration of the AgentCard Smart Contract and the respective fields. Shortened version not listing all variables and functions.}
    \label{fig:SC-AgentCard}
\end{figure}

The agent smart contract serves as a decentralized, blockchain-based identity registry, transforming the conventional A2A AgentCard from a simple and siloed JSON file into a publicly accessible and verifiable smart contract deployed on the blockchain. \autoref{fig:SC-AgentCard} illustrates a shortened version of the Agent Smart Contract. The contract encapsulates all essential agent metadata defined in the A2A AgentCard specification, including name, description, capabilities, endpoints, and skills. Additional contract fields include the contract owner’s address for access control, activity status, and functionality that enables reputation derivation based on transaction frequency and recurring business relationships. Beyond metadata management, the contract enables economic transactions by supporting the receipt of native blockchain coins or approved ERC-20 tokens (such as USDC), and restricts withdrawal permissions to authorized owners. This implementation leverages the AgentExtension object within the AgentCapabilities object of the AgentCard to define a new extension that contains all necessary information for issuing micropayments for agent services.
The extensions field offers a standardized mechanism for extending the core A2A AgentCard specification with additional protocol capabilities. In this context, it enables the direct integration of x402 micropayment information into the advertised capabilities. Each extension follows a structured format and is identified by a unique URI (e.g., urn:a2a-blockchain-x402:extensions:x402:v1), which ensures clear namespace separation and facilitates version management. The extension specifies all necessary fields for an x402 payment, such as asset, network, amount, and payTo, among others. This architecture demonstrates how the A2A protocol can be systematically extended to support economic interactions while maintaining backward compatibility. Agents that do not recognize the x402 extension can still parse the basic AgentCard, whereas payment-enabled clients can automatically initiate appropriate payment workflows based on the advertised parameters.

This approach offers several advantages over traditional AgentCard hosting. It publicly binds agent attributes in an immutable and tamper-proof manner, while also associating each agent with a verifiable payment address. The system further enables transparent payment histories, supporting trustless reputation assessment, and facilitates decentralized discovery, thereby eliminating single points of failure and protect agents against impersonation. Together, these properties lay the groundwork for secure, trustworthy, and interoperable agent identities, enabling autonomous agent economies to operate reliably across organizational boundaries.

\subsection{End-to-End Agent-to-Agent Payment and Settlement Flow}
\begin{figure}
    \centering
    \includegraphics[width=1.1\textwidth]{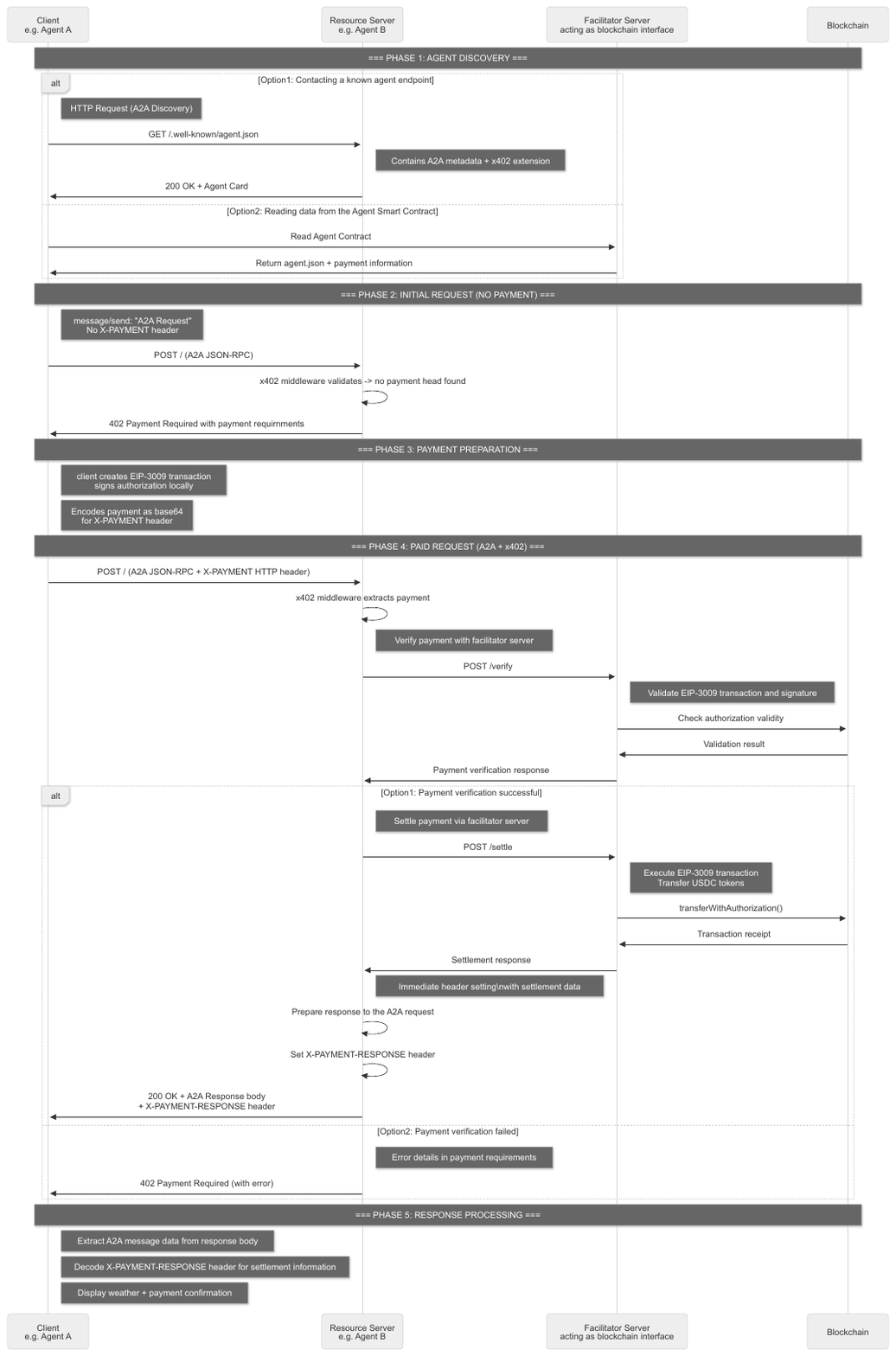}
    \caption{Complete sequence of a paid A2A message, including decentralized agent discovery, micropayment preparation and settlement via x402, and delivery of the agent response.}
    \label{fig:x402+A2A+DLT-full-sequence}
\end{figure}
\autoref{fig:x402+A2A+DLT-full-sequence} illustrates the comprehensive flow for A2A interaction, integrating decentralized agent discovery, micropayment execution using the x402 protocol, and the final delivery of the requested service or data.
The process begins with agent discovery (Phase 1), where a client agent (Agent A) identifies a target service agent (Agent B) by querying either a known endpoint or reading metadata from the agent smart contract on the blockchain. The contract can be discovered as discussed in \ref{sec:sub_agent-discovery}. This metadata includes essential information such as the agent’s skills, endpoint URLs, and payment information. Once the client agent is ready to initiate a service request, it sends an A2A JSON-RPC message to the server agent. If the request does not include payment credentials (i.e., no valid X-PAYMENT header), the server responds with a 402 Payment Required message (Phase 2), detailing the payment requirements and providing the necessary metadata for settlement. The client then prepares an EIP-3009-compliant signed transaction (Phase 3), encoding the payment authorization as a base64 payload and embedding it in the X-PAYMENT header of a new A2A request. Upon receiving this paid request (Phase 4), the x402 middleware on the server agent validates the payment header and communicates with a blockchain facilitator to verify the payment transaction with on-chain data. Next, the payment is settled via the facilitator, which submits the transaction to the blockchain network and pays for the transaction fees. Finally, the resource server processes the A2A request and responds with the requested data or service, including settlement details in the X-PAYMENT-RESPONSE header. In the event of a failed payment, the server may issue a 402 response with error details. The client can then decode the X-PAYMENT-RESPONSE to confirm settlement and process the returned information (Phase 5). 

\section{Discussion}

This work presents a practical approach towards a truly decentralized multi-agent economy by addressing the two most critical gaps in the current Agent2Agent protocol specification: agent discoverability and payment. 

Firstly, by anchoring AgentCards in on-chain smart contracts, agents are provided with a sovereign, publicly available identity. This identity is further enhanced with additional features, including payment details and functionalities that any peer can access. The approach introduces decentralized, publicly accessible, and discoverable agent identities, enabling dynamic updates of agent metadata and establishing a foundation for decentralized reputation systems based on transaction histories. Secondly, by seamlessly integrating the x402 micropayment flow into the A2A protocol through standard HTTP headers, frictionless, blockchain-agnostic payments are facilitated between agents. The implemented prototype utilizes USDC tokens and the EIP-3009 standard on the Ethereum blockchain. However, it is compatible with various Blockchains and tokens. The additions evolve A2A from a pure messaging protocol into a protocol that enables autonomous multi-agent economies..

Despite this progress, key challenges remain. Although the presented design presents and discusses multiple discovery mechanisms, including factory contracts, both permissioned and permissionless on-chain registries, off-chain indexers, and emerging ERC-based standards. However, there is still a lack of unified registries that span the entire ecosystem, which makes peer discovery in large or fragmented networks challenging. While x402 payments are lightweight, embedding on-chain settlements into latency-sensitive workflows can introduce unpredictable delays and fee variability, which may require mitigation via dedicated payment channels or pre-funded state channels. Allowing agents to access private keys and funds opens up new attack vectors, such as the risk of attackers attempting to drain the funds of the agent. For instance, an attacker could persuade an agent to authorize a large payment. Future research should focus on investigating these potential attack vectors and developing strategies to protect against such attacks.

Achieving truly autonomous multi-agent economies requires comprehensive standardization efforts that extend beyond current protocol specifications. While this work demonstrates technical feasibility, widespread adoption and interoperability necessitate addressing several critical standardization challenges. A foundational step is the development of a standardized ERC (Ethereum standard) specification for AgentCards and agent identities. Such a standard would unify on-chain agent identities across platforms, enabling blockchain explorers, indexers, and aggregation services to natively catalog and discover service agents. It would further allow developers to access agents through a predictable interface, significantly contributing to the realization of a multi-agent economy. This standardization could mirror the success of token standards, such as ERC-20 and ERC-721, which have driven ecosystem growth through community-driven aggregation and cross-platform compatibility. The proposed ERC specification should define a minimal yet comprehensive set of variables and functions for smart contracts to establish verifiable on-chain agent identities. Core requirements include standardized metadata fields for agent capabilities, service endpoints, payment preferences, and reputation metrics, as well as extensibility mechanisms to support future enhancements and maintain backward compatibility. 

In addition, certification programs for agents can provide an extra layer of assurance and quality control. Such programs, governed by industry consortiums or DAOs, could include rigorous testing of functional compliance, security best practices, and ethical guidelines. Successfully certified agents would earn on-chain credentials that signal verified competence and adherence to standards. These credentials, combined with reputation scores, offer users and other agents a clear signal of trustworthiness, reducing onboarding friction and fraud risk.

Another important requirement is the establishment of a machine-interpretable ontology to represent the skills and capabilities of advertising agents. By formalizing a semantic layer that provides a shared, structured vocabulary for service description, agents can articulate their offerings in a standardized manner. This semantic framework facilitates automated interpretation, comparison, and matching of services. Consequently, this approach enhances interoperability and enables more sophisticated, automated negotiation and coordination among heterogeneous agents across diverse platforms.

Enabling agents to make complex economic decisions autonomously is a fundamental research frontier. Decision-making frameworks must extend beyond price comparison to encompass service quality, provider reputation, response time, reliability, and increasingly sophisticated payment models. For example, agents should be able to recognize and negotiate volume-based discounts, subscription or membership tiers, loyalty rewards, and dynamic pricing offers that adjust in real time based on demand or market conditions. An agent might commit to a flat-rate subscription for ongoing access to a suite of services, securing predictable budgeting and priority support, or it might automatically apply bulk-purchase discounts when scheduling repeated interactions. 

As agent autonomy increases, the role of human oversight becomes more nuanced. Adaptive human-in-the-loop mechanisms should intelligently escalate decisions to users when transactions involve high value, novel contract terms (such as bespoke subscription commitments), or heightened uncertainty around performance guarantees. User interfaces must present clear decision support across all payment models, combining cost-benefit analyses, projected savings from discounts or subscription plans, ongoing risk assessments, and confidence levels for each recommendation, to ensure that end users retain visibility and control over critical economic choices.

In summary, by combining A2A messaging, on-chain discovery, and HTTP 402 micropayments into a cohesive framework, this architecture marks a significant advancement toward creating an open and scalable web of autonomous service agents. By providing agents with immutable identities, discoverable endpoints, and built-in economic flows, this approach makes considerable progress toward realizing an open, scalable web of autonomous service agents. Ongoing efforts in standardization, along with the development of ecosystems for discovery, reputation, and governance, will be essential for transitioning from prototype to production and fully unlocking the potential of multi-agent economies.
\section{Conclusion}
This paper introduces a novel architecture that advances the A2A protocol by enabling decentralized agent discoverability and seamless micropayments. By implementing AgentCards as on-chain smart contracts, agents are provided with immutable, publicly accessible identities that support dynamic updates, payment, and facilitate the development of decentralized reputation systems based on transaction history. The integration of the x402 micropayment protocol into A2A, utilizing standard HTTP headers, enables frictionless, blockchain-agnostic payments between agents, transforming A2A into a robust economic protocol. This approach supports economically viable collaboration among autonomous agents operating across organizational boundaries.

The prototype demonstrates practical feasibility, utilizing USDC and EIP-3009 on Ethereum, while maintaining compatibility with other EVM chains and tokens. Multiple discovery mechanisms are supported, enabling a flexible and composite discovery strategy. Nevertheless, challenges remain, such as the absence of a unified, ecosystem-wide registry and the latency introduced by on-chain settlements.

Future work should focus on standardizing ERC specifications for AgentCards, developing decentralized and permissioned registries, and embedding reputation or staking mechanisms to further incentivize quality service. By merging A2A messaging, distributed ledger-backed discovery, and HTTP-402 micropayments, this architecture lays the groundwork for an open, scalable web of autonomous service agents that can discover, authenticate, and transact with minimal human intervention. Continued standardization and ecosystem collaboration will be crucial for moving from prototype to widespread adoption and realizing the full potential of multi-agent economies.

%
%
%
\bibliographystyle{splncs04}
\bibliography{references,references_manual}
\end{document}